\documentclass[epj,referee]{svjour}

\usepackage{graphics}
\usepackage{amsmath}
\newcommand{\e}{\mbox{\large\em e}}

\begin{document}

\title{Lifetime distributions in the methods of non-equilibrium statistical operator and superstatistics}

\author{V.V. Ryazanov}

\institute{Institute for Nuclear Research, Kiev, pr.Nauki, 47
Ukraine; \email{vryazan@kinr.kiev.ua}}
\date{Received: date / Revised version: date}
%

\abstract{A family of non-equilibrium statistical operators
is introduced which differ by the system age distribution
over which the quasi-equilibrium (relevant) distribution is
averaged. To describe the nonequilibrium states of a system we introduce a
new thermodynamic parameter - the lifetime of a system.
Superstatistics, introduced in works of Beck and Cohen [Physica A \textbf{322}, (2003), 267] as
fluctuating quantities of intensive thermodynamical parameters,
are obtained from the statistical distribution of lifetime (random
time to the system degeneracy) considered as a thermodynamical parameter. It
is suggested to set the mixing distribution of the fluctuating
parameter in the superstatistics theory  in the form of the
piecewise continuous functions. The distribution of lifetime in
such systems has different form on the different stages of
evolution of the system. The account of the past stages of the evolution of a system
 can have a substantial impact on the non-equilibrium behaviour of the system
in a present time moment.}

%
\maketitle

\section{Introduction}
\label{intro}

One of the most promising ways of development of the description
of the non-equilibrium phenomena is provided by the method of the
non-equilibrium statistical operator (\emph{NSO})
\cite{zub71,zub80,zub96}. In \cite{ry01} a new interpretation of
the \emph{NSO} method is given, where the operation of taking the
invariant part \cite{zub71} or the use of the auxiliary "weight
function" (in the terminology of \cite{ra95,ra99}) in \emph{NSO}
are treated as averaging of the quasi-equilibrium statistical
operator over the distribution of the past lifetime span (age) of
the system. In \cite{ra95,ra99} it is noted, that multiple choice
of the "weight functions" can be taken. In \cite{zub80} a  uniform
distribution over an initial moment $t_{0}$ is considered, which
after the change of integration order reduces to the exponentially
distributed weight function $p_{q}(u)= \exp\{-\varepsilon
u\}$ in (\ref{NSO}). Such distribution is the limiting case of the
lifetime distribution \cite{str61}, that is the distribution of
the first passage time of a given level. The term "lifetime"
denoting the time of the first passage of a level was used in
\cite{str61}. Encountered in the literature is also the term
"non-equilibrium relaxation time", as well as some others.

The form of the function $p_{q}(u)= \exp\{-\varepsilon u\}$
in (\ref{NSO}) is connected with the form of the source in the
Liouville equation for \emph{NSO}. In \cite{rep83,der85} the
sources in the Liouville equation different from that introduced
in the \emph{NSO} method in \cite{zub71,zub80} are considered. The
approach of the present paper differs from the methods used in
\cite{rep83,der85}. But the use of the distribution of the system
lifetime in the present work can be compared with that in
\cite{rep83} enlarging the set of macroobservables;
besides the common physical macroobservables this approach
includes additional ones, namely the life span (lifetime).

In \cite{luz} other alternative derivations of
\emph{NSO} are performed, following the ideas proposed by McLennan  \cite{mc},
and a relation with an earlier proposal by I. Prigogine
\cite{Prig} is discussed. The source in the Liouville equation can
in principle take different forms. The form of a source used in
\cite{zub71,zub80,zub96} is a specific case which can be obtained under
the assumption of the weak coupling limit of the interaction of the system
with its environment.

In \cite{ry07} it is shown, what is the impact of changing the
system lifetime distribution to the non-equilibrium properties of
system with finite volume. In the present work we consider
infinitely big systems with infinite average lifetimes as well.

In \cite{zub80} a physical interpretation of the exponential
distribution for $p_{q}(u)$ is given: the system freely evolves
as an isolated system governed by the operator of
Liouville. In addition, the system undergoes random transitions at
which its phase point representing the system spreads from one
phase trajectory to another one in a random fashon with an exponential probability
under the influence of a "thermostat", thus average intervals
between random jumps increase infinitely. This feature reflects in
the parameter of the exponential distribution tending to zero after the
thermodynamic limiting transition. Real physical systems have
finite sizes. The exponential distribution describes completely
random systems. The influence of the surrounding on a system can
have organised character as well, for example, for the systems
in the non-equilibrium steady-states with input and output flows.
The ways of the interaction of the system with surrounding can be
different, therefore various choices of the functions $p_{q}(u)$
are justified.

Nonextensive statistical mechanics \cite{ts,abe}  can be regarded
as an embedding of the common statistical mechanics into a more
general framework.  Many complex systems exhibit a
spatio-temporally inhomogeneous dynamics that can be effectively
described by a superposition of several statistics on different
time scales, termed "superstatistics" \cite{beck,beck1}.
Nonequilibrium situations are described by a fluctuating parameter
$\beta$, which can be, for example, the inverse temperature. The
generalization of the Boltzmann factor $\exp\{-\beta E\}$
was introduced in the following form:
\begin{equation}
B(E)=
\int_{0}^{\infty}\mathrm{d}\beta'\,f(\beta')\exp\{-\beta'
E\}. \label{su}
\end{equation}

The type of superstatistics induced depends on the
probability distribution $f(\beta)$ of the parameter $\beta$. The
special case of these superstatistics, with the function $f$ in the form of
gamma-distribution, appears in the nonextensive statistical
mechanics \cite{beck,beck1,ts,abe}, describing a number of
physical phenomena which are not satisfactory described by the
Boltzmann-Gibbs statistics. In the present work the
superstatistics like (\ref{su}) (together with its generalization)
is obtained starting from the distribution which contains a
lifetime of a statistical system as a thermodynamic variable
\cite{ry04,rys04,ry05,rys06}, Section 5. This distribution has been
applied earlier to the description of aerosol behaviour \cite{ryaer}, and
neutrons in a nuclear reactor \cite{ryat}.

In several works, for example in \cite{beck,beck1}, the
distribution function $f(\beta)$ is introduced as some continuous
function expressed through arbitrary analytical form of the
distribution of a random variable $\beta$. But the definition of the
continuous density of distribution assumes its piecewise
continuous character when the density of distribution has finite
number of breaks. Real nonequilibrium systems, as a rule, are
spatially non-uniform. This behaviour can be mathematically
described by the piecewise continuous functions, the examples
of which are given in the present work.

\section{Modifications to the nonequilibrium description}
\label{sect:2}

In \cite{ry01} a new interpretation of the method of the
\emph{NSO} is given. Let us consider now, what consequences follow
from such interpretation of \emph{NSO}. Setting various
distributions for past lifetime of the system $p_{q}(u)$, we
obtain a family of \emph{NSO}, where the exponential distribution
in Zubarev \emph{NSO} is a particular choice of the form of the (arbitrary) lifetime
distribution $p_{q}(u)$. The class of \emph{NSO} from this family
can be related to the class of the lifetime (or age) distributions
(taken, for example, from the stochastic theory of storage
processes, the theory of queues etc) and to the relaxation
properties of that class of physical systems which is
investigated. The general expression for \emph{NSO} with an arbitrary
distribution $p_{q}(u)$ is
\begin{align}
&\ln\varrho(t)=
\int_{0}^{\infty}p_{q}(u)\ln\varrho_{q}(t-u,
-u)\,\mathrm{d}u=
\label{NSO} \\
&=\ln\varrho_{q}(t, 0)-\int\limits_{0}^{\infty}\left(\int
p_{q}(u)\,\mathrm{d} u\right) \frac{\partial \ln\varrho(t-u,
-u)}{\,\partial u}\,\mathrm{d}u\,, \nonumber
\end{align}
\[
\ln\varrho_{q}(t, 0)=-\Phi(t)-\sum_{n}F_{n}(t)P_{n};
\]
\[
\ln\varrho_{q}(t,
t_{1})=\e^{\textstyle\left\{-t_{1}H/i\hbar\right\}}\ln\varrho_{q}(t,
0)\e^{\textstyle\left\{t_{1}H/i\hbar\right\}};
\]
\[
\Phi(t)=\ln\,\mathrm{Sp}\exp\{-\sum_{n}F_{n}(t)P_{n}\}\,,
\]
where $H$ is Hamiltonian, $\ln\varrho(t)$ is the logarithm of the
\emph{NSO}, $\ln\varrho_{q}(t, 0)$ is the logarithm of the
quasi-equilibrium (or relevant) distribution; the first time
argument indicates the time dependence of the values of the
thermodynamic parameters $F_{m}$; the second time argument $t_{2}$
in $\varrho_{q}(t_{1}, t_{2})$ denotes the time dependence through
the Heizenberg representation for dynamical variables $P_{m}$ on
which $\varrho_{q}(t, 0)$ can depend
\cite{zub71,zub80,zub96,ra95,ra99}, integration by parts in time
is carried out at  $\int p_{q} (y)\,\mathrm{d}y_{|y=0}=-1; \int
p_{q}(y)\,\mathrm{d}y_{|y\rightarrow\infty}=0$. If
$p_{q}(y)=\varepsilon\exp\{-\varepsilon y\};
\varepsilon=1/\langle\Gamma\rangle$, then the expression
(\ref{NSO}) passes in \emph{NSO} from \cite{zub71,zub80}. In
\cite{ry01} the auxiliary weight function \cite{ra95,ra99}
$p_{q}(u)=\varepsilon \exp\{-\varepsilon u\}$ was interpreted as
the density of the probability distribution of the lifetime
$\Gamma$ of a system. There $\Gamma$ is a random variables of the
lifetime of a system from the moment $t_{0}$ of its birth till the
current moment $t$; $\varepsilon^{-1}=\langle t-t_{0}\rangle$;
$\langle t-t_{0}\rangle=\langle\Gamma\rangle$; $u=t-t_{0}$. This
value represents the age of system. The operator of internal time
describing the age of a system was also introduced in
I.Prigogine's works (for example, see \cite{Prig}). If the
interval $t-t_{0}=u$ is large enough (that is the details of an
initial condition as dependence on the initial moment $t_{0}$ are
nonsignificant and nonphysical \cite{zub71,zub80}), it is possible
to introduce the minimal lifetime $\Gamma_{min}=\Gamma_{1}$ and to
integrate in (\ref{NSO}) over the interval $(\Gamma_{1},\infty)$.
It is possible to specify many concrete expressions for lifetime
distribution of a system, each of which possesses its own
advantages. Each of these expressions induces some form of a
source in the Liouville equation for the nonequilibrium
statistical operator. In the general case of an arbitrary function
$p_{q}(u)$ the source is:
\begin{equation}
J=p_{q}(0)\ln\varrho_{q}(t,
0)+\int_{0}^{\infty}\frac{\partial p_{q}(y)}{\partial
y}\Big(\ln\varrho_{q}(t-y, -y)\Big)\,\mathrm{d}y
\label{sour}
\end{equation}
(if the value $p_{q}(0)$ diverges, it is necessary to choose the
lower limit of integration equal to some $\Gamma_{min}>0$). Such
approach corresponds to the dynamic Liouville equation in the form
of Boltzmann-Bogoliubov-Prigogine \cite{ra95,ra99,Prig},
containing dissipative items. In \cite{mor} it was noted that the
role of the form of the source term in the Liouville equation in
\emph{NSO} method has never been investigated.

Let us notice, that in the case when the value
\\ $\partial\ln\varrho_{q}(t-y,-y)/\,\partial y$ (the operator of entropy production
$\sigma$ \cite{zub71,zub80}) in the second term of the r.h.s.
(\ref{NSO}) does not depend on $y$ and can be taken out from the
integration on $y$, this second term takes on the form
$\sigma\langle\Gamma\rangle$, and the expression (\ref{NSO}) thus
does not depend on the form of the function $p_{q}(y)$. It is the
case, for example, if $\varrho_{q}(t)\sim \exp\{-\sigma t\},
\sigma=\mathrm{\emph{const}}$. In \cite{dew} such a distribution
is obtained from the principle of maximum of entropy applied to
the set of average values of fluxes.

As it is known (for example, \cite{str61}), the
exponential distribution for lifetime
\begin{equation}
p_{q}(y) =\varepsilon\exp\{-\varepsilon  y\}, \label{expDi}
\end{equation}
used in the works of Zubarev \cite{zub71,zub80}, is the limiting
distribution for the lifetime, valid for large times. Thus, in
\cite{zub71,zub80} the thermodynamic results are obtained, which in
this context are valid for all systems.

For the \emph{NSO} (\ref{NSO}) with the function (\ref{expDi}) in
the form suggested by Zubarev the value in the second item is
\begin{align}
-\int p_{q}(y)\,\mathrm{d}y =\exp\{-\varepsilon
y\}=1-\varepsilon
y+(\varepsilon y)^{2}/2-...= \label{sum} \\
1-y/\langle\Gamma\rangle+y^{2}/2\langle\Gamma\rangle^{2}-\dots .
\nonumber
\end{align}
Evidently the average lifetime tends to infinity,
$\langle\Gamma\rangle\rightarrow\infty$, and the correlation (\ref{sum})
tends to unity.

Besides the exponential density of probability (\ref{expDi}),  the Erlang distributions (special or
 general form), gamma distributions etc (see \cite{co,cox}), as well as the
modifications considering subsequent composed asymptotic
of the decomposition \cite{tur} can be used as
candidates for the density of lifetime distribution. Gamma distributions
describe the systems whose evolution acquires some stages (number of
these stages is given by the order of gamma distribution). Considering
actual stages in non-equilibrium systems (chaotic, kinetic,
hydrodynamic, diffusive and so forth \cite{bog}), it
is possible to justify the use of gamma distributions of a
kind
\begin{equation}
p_{q}(y) =\varepsilon(\varepsilon y)^{k-1}\exp\{-\varepsilon
y\}/\Gamma(k) \label{gam}
\end{equation}
($\Gamma(k)$ is gamma function, passing at $k=1$ to exponential one
(\ref{expDi})), and to understand their
importance in the description of non-equilibrium properties. The
piecewise-continuous distributions corresponding to different
stages of evolution of the system will be used below.

More accurate specifying the shape of the function $p_{q}(u)$ in
comparison with limiting exponential function (\ref{expDi}) allows
to describe in more detail the real stages of evolution of a
system. Every form of the lifetime distributions has certain
physical sense. In \cite{ry07} additional terms to \emph{NSO} in
Zubarev form for the gamma distribution (\ref{gam}) are obtained.

\section{Systems with infinite lifetime}
 \label{sect:3}

The amendments to \emph{NSO} in
the Zubarev form in the Section 2 and in \cite{ry07} are obtained for the systems of finite sizes
and lifetimes. We will show further, how the same effects, involving
the influence of the past of system on its present non-equilibrium
state have impact on the systems with infinite lifetime, for
example, for the systems of infinite volume after thermodynamic
limiting transition.

Amendments to the unity term in the equation (\ref{sum}) in \cite{ry07}
become vanishingly small when the size of the system and its
average lifetime tend to infinity, as in the model distribution
(\ref{expDi}) used in Zubarev \emph{NSO}. For the systems of
finite size but still exponential distribution these terms result in
nonzero amendments to the expression (\ref{sum}). Thus, these
additional terms to \emph{NSO} and hence to the kinetic equations, kinetic
coefficients and other non-equilibrium characteristics of the
system are in fact the impact of the finiteness of size and
lifetime of the system. They do not result merely from the choice
of particular form of the lifetime distribution in
the system. In what follows we investigate whether there exist such
distributions of the lifetime of system for which the additional
contribution to \emph{NSO} differs from Zubarev \emph{NSO} even
for infinitely large systems with infinitely large lifetime.

We will consider several examples of choosing the function $p_{q}(u)$ in
(\ref{NSO}). We limit ourselves to the piecewise-continuous
distributions, from where the results different from
\cite{zub71,zub80} follow. There are numerous experimental
evidences of such changes of the distribution of lifetime of the
system $p_{q}(u)$ on the time scale of the life span of the
system. In \cite{inoue,mantegna} the transition of the
distribution of the first passage processes from Gaussian regime
to the non-L\'{e}vy behaviour in a specific time moment is shown.
Real systems possess finite sizes and finite lifetime which
implies the influence of surrounding on them. The fact that these
sources do not vanish in a limit of infinitely large systems, is
related to the openness of the system, hence to the influence of
its surrounding.

\textit{3a)} We shall set
\begin{equation}
p_{q}(u)= \left\{
\begin{array}{ll}
{\displaystyle \frac{ka^{k}}{(u+a)^{k+1}}}, & \quad u<c\,;
\\
b \varepsilon\exp\{-\varepsilon u\}, & \quad u\geq c\,.
\end{array}
\right.
\label{pa}
\end{equation}
From the condition of the normalization
$\int_{0}^{\infty}p_{q}(u)=1$ the normalization
multiplier is $b={\displaystyle \e^{\varepsilon c}\left(\frac{a}{a+c}\right)^{k}}$. Average
lifetime $\langle\Gamma\rangle$ for the distribution (\ref{pa}) is
equal to
\begin{equation}
\langle\Gamma\rangle=\frac{a}{k-1}+\left(\frac{a}{a+c}\right)^{k}\left[\frac{1}{\varepsilon}(1+\varepsilon
c)-\frac{(kc+a)}{(k-1)}\right]. \label{ava}
\end{equation}
The value (\ref{ava}) $\langle\Gamma\rangle \rightarrow \infty$ at
$\varepsilon\rightarrow0$. From (\ref{pa}) we find
\begin{equation}
-\int p_{q}(u)=\left\{
\begin{array}{ll}
{\displaystyle \left(\frac{a}{u+a}\right)^{k}}\,, & \quad u<c;
\\
b \exp\{-\varepsilon u\}\,, & \quad u\geq c.
\end{array}
\right. \label{pa1}
\end{equation}

The source in the  right part of Liouville equation for the
distribution (\ref{pa}) in accordance with the expression (\ref{sour})
equals
\begin{align*}
J=\frac{k}{a}
\ln\varrho_{q}(t,0)-\int_{0}^{c}{\displaystyle \frac{k(k+1)a^{k}}{(u+a)^{k+2}}}S\,\mathrm{d}u-\\
\e^{
c\varepsilon}\left(\frac{a}{a+c}\right)^{k}\int_{c}^{\infty}\varepsilon^{2}\e^{-\varepsilon
u}S\,\mathrm{d}u,
\end{align*}
where $S=\ln\varrho_{q}(t-u,-u)$. The distribution of
\emph{NSO} (\ref{NSO}) in the case of (\ref{pa}) is
\[
\ln\varrho(t)=\ln_{zub}\varrho(t)+\Delta;
\]
\begin{align*}
\Delta=\int_{0}^{c}\left[\left(\frac{a}{a+u}\right)^{k}-\e^{-\varepsilon u}\right]\sigma\,\mathrm{d}u+ \\
\int_{c}^{\infty}\left[\e^{\varepsilon
c}\left(\frac{a}{a+c}\right)^{k}-1\right]\e^{-\varepsilon u}\sigma\, \mathrm{d}u,
\end{align*}
where $\sigma=\partial \ln\varrho(t-u, -u)/\partial u$;
$\ln_{zub}\varrho(t)=\ln\varrho_{q}(t, 0)+\\
\int_{0}^{\infty}\e^{-\varepsilon u}\sigma\,\mathrm{d}u$ is the
distribution obtained by Zubarev in \cite{zub71,zub80}, and
$\Delta$ is a finite amendment to it.

\textit{3b)} We will consider now the distribution of the following kind:
\begin{equation}
p_{q}(u)=\left\{
\begin{array}{ll}
d\,, & \quad u<c;
\\
\varepsilon \exp\{-\varepsilon u\}\,, & \quad u\geq c .
\end{array}
\right.
\label{pqb}
\end{equation}
From the condition of the normalization we find \\
$d={\displaystyle\frac{1}{c}}\left(1-\e^{-\varepsilon c}\right)$. The average lifetime is
\begin{equation}
\langle\Gamma\rangle=\frac{dc^{2}}{2}+\frac{1}{\varepsilon}\e^{-\varepsilon
c}(1+\varepsilon c)\,.\label{avb}
\end{equation}
The average lifetime $\langle\Gamma\rangle\rightarrow\infty$ at
$\varepsilon\rightarrow 0$. The source (\ref{sour}) in the Liouville
equation in this case equals
\[
J=d
\ln\varrho_{q}(t,0)-\int_{c}^{\infty}\varepsilon^{2}\e^{-\varepsilon
u}S\,\mathrm{d}u.
\]
The amendment to the Zubarev form of \emph{NSO} is
\[
\Delta=-\int_{0}^{c}\left[\e^{-\varepsilon
u}+\frac{1}{c}\left(1-\e^{\textstyle -\varepsilon c}\right)u\right]\sigma \,\mathrm{d}u.
\]

We see in this case, that an additional memory term of the
system stems from the finiteness of its size, and the limited memory effect is observed.
It is possible to consider other examples of the functions
$p_{q}(u)$ which result in the limited memory effects.

\textit{3c)} For the exponential density of distribution but with different
intensities in different time intervals
\begin{equation}
p_{q}(u)=\left\{
\begin{array}{ll}
\varepsilon_{1} \exp\{-\varepsilon_{1} u\}\,, & \quad u<c\,;
\\
b\varepsilon_{2} \exp\{-\varepsilon_{2} u\}\,, & \quad u\geq c\,,
\end{array}
\right. \label{pqc}
\end{equation}
from the condition of the normalization it follows that
$b=\e^{c(\varepsilon_{2}-\varepsilon_{1})}$;
\begin{equation}
\langle\Gamma\rangle=\frac{1}{\varepsilon_{1}}\left[1-\e^{-\varepsilon_{1}
c}(1+\varepsilon_{1}
c)\right]+\frac{1}{\varepsilon_{2}}\e^{-\varepsilon_{1}
c}(1+\varepsilon_{2} c) \,.  \label{avc}
\end{equation}
At $\varepsilon_{2}\rightarrow 0$,
$\langle\Gamma\rangle\rightarrow\infty$.
\begin{align*}
J=\varepsilon_{1}
\ln\varrho_{q}(t,0)-\int_{0}^{c}\varepsilon_{1}^{2}\e^{-\varepsilon_{1}
u}S\,\mathrm{d}u- \\
\e^{c(\varepsilon_{2}-\varepsilon_{1})}\int_{c}^{\infty}\varepsilon_{2}^{2}\e^{-\varepsilon_{2}
u}S\,\mathrm{d}u\,;
\end{align*}
\[
\Delta=\int\limits_{0}^{c}\left[\e^{-\varepsilon_{1} u}-\e^{-\varepsilon_{2}
u}\right]\sigma
\,\mathrm{d}u+\int\limits_{c}^{\infty}\left[\e^{c(\varepsilon_{2}-\varepsilon_{1})}-1\right]\e^{-\varepsilon_{2}
u}\sigma \,\mathrm{d}u\,;
\]
\[
\Delta_{\varepsilon_{2}\rightarrow 0}\rightarrow
\int_{0}^{\infty}\left[\e^{-\varepsilon_{1} u}-1\right] \sigma \,\mathrm{d}u\,.
\]

The natural question is now why do the examples of this section
differ from the examples of Section 2. In the interpretation of
\cite{zub80} it is  the random value $t_{0}$ in $u=t-t_{0}$ that
fluctuates. In \cite{zub80} the limiting transition is performed
for the parameter $\varepsilon, \varepsilon\rightarrow 0$ in the
exponential distribution $p_{q}(u)=\varepsilon
\exp\{-\varepsilon u\}$ after passing to the thermodynamic
limit. In the interpretation of \cite{ry01} this corresponds to
the fact that the mean lifetime of the system
$\langle\Gamma\rangle=\langle
t-t_{0}\rangle=1/\varepsilon\rightarrow\infty$. But average
intervals between random jumps infinitely incease, exceeding the
lifetime of the system. Therefore a source term in the Liouville
equation tends to zero. If the change of the distribution
$p_{q}(u)$ caused by the influence of the surrounding, occurs on
the time interval of the life span of the system, as in examples
3a)-3c), its impact remains even if the mean lifetime tends to
infinity.

\section{Application of the distributions of Section 3 to the conductivity}
 \label{sect:4}

On the example of conductivity we will investigate,
what are consequences of the change of the type of functions $p_{q}(u)$
and $\varrho(t)$ as compared to the exponential law for
$p_{q}(u)$, used in \cite{zub96}.

The determination of the conductivity coefficient by the \emph{NSO}
method is considered in \cite{kal74,rep81,rep91a,bobr}. In
this section we will describe the transport of charges in the
electric field, as linear reaction on a mechanical perturbation,
that is we regard the electric conductivity in the linear approximation, following
the results of \cite{zub96} and, as in \cite{zub96}, we
limit ourselves to the important special case -- the reaction of
the equilibrium system to the spatially homogeneous variable field
\[
\vec{E}^{0}(t)=\int_{-\infty}^{\infty}\frac{d\omega}{2\pi}\e^{-i\omega
t}\tilde{\vec{E}}^{0}(\omega).
\]
The Hamiltonian of perturbation is given by
\[
H^{1}_{t}=-\vec{P}\vec{E}^{0}(t),
\]
where $\vec{P}$ is the operator corresponding to the vector of
polarization of the system. In the coordinate representation this
operator is written as
\[
\vec{P}=\sum_{i}e_{i}\vec{r}_{i},
\]
where $e_{i}$ is the charge of the particle, and $\vec{r}_{i}$ is its
position vector. The operator of current is
\[
\vec{J}=\dot{\vec{P}}=e\sum_{j}\dot{\vec{r}}_{j}=\frac{e}{m}\sum_{j}\vec{p}_{j},
\]
where $\vec{p}_{j}$ is particle momentum, $m$ is mass of a
particle. We choose a model in which the Coulomb interaction is
taken into account as a self-consistent screening of the field, i.e. we
take $\vec{E}=\vec{E}^{0}$. The most essential
difference from \cite{zub96} consists in the replacement of the Laplace transformation used in
\cite{zub96}, that is
\begin{equation}
\langle A;B \rangle_{\omega+i\varepsilon}=
\int_{0}^{\infty}\mathrm{d}t\,\e^{i(\omega+i\varepsilon)t}(A(t),B)),\,
(\varepsilon>0)\,, \label{la}
\end{equation}
where $(A(t),B(t'))=\int_{0}^{1}dx\mathrm{Tr}\{\Delta
A(t)\Delta B(t'+i\beta \hbar x)\varrho_{eq}\}$ is the time
correlation function \cite{zub96}, by the another integral
transformation. So, for the example \textit{3a)} with the
distribution of the form (\ref{pa}), (\ref{pa1}) the expression
(\ref{la}) is replaced by
\begin{equation}
\langle A;B \rangle_{\omega;a,k}+\mathrm{e}^{\varepsilon
c}{\displaystyle \left(\frac{a}{a+c}\right)^{k}}\langle A;B
\rangle_{\omega+i\varepsilon;(c,\infty)}, \label{la1}
\end{equation}
where
\[
\langle A;B \rangle_{\omega;a,k}=\int_{0}^{c}dt\,\e^{i\omega
t}{\displaystyle\left(\frac{a}{a+t}\right)^{k}}(A(t),B);
\]
\begin{equation}
\langle A;B
\rangle_{\omega+i\varepsilon;(c,\infty)}=\int_{c}^{\infty}dt\,\e^{i(\omega+i\varepsilon)t}(A(t),B)\,.\label{nla}
\end{equation}

We consider an isotropic environment in which the tensor of
conductivity is diagonal. In \cite{zub96} the expression for the Laplace
transform of the kind (\ref{la}) for the specific
resistance $\rho(\omega)$ is obtaned:
\begin{equation}
\rho(\omega)=\frac{1}{\sigma(\omega)}=\frac{3V}{\beta(\vec{J},\vec{J})}[-i\omega+M];
\label{re1}
\end{equation}
\begin{equation}
M=\frac{\langle\dot{\vec{J}};\dot{\vec{J}}\rangle_{\omega+i\varepsilon}}
{(\vec{J},\vec{J})+\langle\dot{\vec{J}};\vec{J}\rangle_{\omega+i\varepsilon}},
\label{m}
\end{equation}
where $V$ is the volume of the system, $\beta$ is inverse
temperature, $\sigma(\omega)$ is the scalar coefficient of
conductivity. In the examples considered
below in expressions (\ref{re1})-(\ref{m}) it is the value $M$ that changes.
Performing the operations of \cite{zub96}, with
replacement of expression (\ref{la}) by (\ref{la1}), in place of correlation (\ref{m}) we obtain a more
complicated expression of the kind
\[
M=\frac{\langle\dot{\vec{J}};\dot{\vec{J}}\rangle_{\omega;a,k}+
{\displaystyle\frac{i\omega}{i(\omega+i\varepsilon)}}{\textstyle\e}^{\textstyle\varepsilon
c}{\displaystyle\left(\frac{a}{a+c}\right)^{k}}\langle\dot{\vec{J}};\dot{\vec{J}}\rangle_{\omega+i\varepsilon;(c,\infty)}}
{K};
\]
\begin{align*}
& K=(\vec{J}(0),\vec{J})-\frac{k}{a}\langle\vec{J};\vec{J}\rangle_{\omega;a,k+1}- \\
&\left(1-\frac{i\omega}{i(\omega+i\varepsilon)}\right) \e^{i\omega
c}\left(\frac{a}{a+c}\right)^{k}(\vec{J}(c),\vec{J})+ \\
&+\langle\dot{\vec{J}};\vec{J}\rangle_{\omega;a,k}+\frac{i\omega}{i(\omega+i\varepsilon)}\e^{\varepsilon
c}{\displaystyle\left(\frac{a}{a+c}\right)^{k}}
\langle\dot{\vec{J}};\vec{J}\rangle_{\omega+i\varepsilon;(c,\infty)}\,.
\end{align*}
At $\varepsilon\rightarrow 0$ and
$\langle\Gamma\rangle\rightarrow\infty$,
$K\rightarrow(\vec{J}(0),\vec{J})-{\displaystyle\frac{k}{a}}\langle\vec{J};\vec{J}\rangle_{\omega;a,k+1}+
+ \langle\dot{\vec{J}};\vec{J}\rangle_{\omega;a,k}+
{\displaystyle\left(\frac{a}{a+c}\right)^{k}}\langle\dot{\vec{J}};\vec{J}\rangle_{\omega;(c,\infty)}.
$

For the distribution (\ref{pqb}) in \textit{3b)} the Laplace transform of the kind
(\ref{la}) is replaced by
\[
-d\langle A;B \rangle_{\omega;c,t}+\langle A;B
\rangle_{\omega+i\varepsilon;(c,\infty)},
\]
where $\langle A;B \rangle_{\omega;c,t}=\int_{0}^{c}\mathrm{d}t\, \e^{i\omega
t}t(A(t),B)$, the value $d$ is given in (\ref{pqb}), (\ref{avb}),
$\langle A;B \rangle_{\omega+i\varepsilon;(c,\infty)}$ is given in
(\ref{nla}). Instead of (\ref{m}) in this case we will
get the expression
\[
M=\frac{M_{1}}{K_{1}} \,,
\]

\begin{align*}
M_{1}=\frac{i\omega}{i(\omega+i\varepsilon)}\langle\dot{\vec{J}};
\dot{\vec{J}}\rangle_{\omega+i\varepsilon;(c,\infty)}- \
\mathrm{d} \left[\langle\dot{\vec{J}
};\dot{\vec{J}}\rangle_{\omega;c,t}+
\langle\vec{J};\dot{\vec{J}}\rangle_{\omega;c,t=1}\right]\,,
\end{align*}
\begin{align*}
K_{1}=\frac{i\omega}{i(\omega+i\varepsilon)}\langle\dot{\vec{J}};\vec{J}\rangle_{\omega+i\varepsilon;(c,\infty)}-
d\left[\langle\dot{\vec{J}};\vec{J}\rangle_{\omega;c,t}+ \right.  \\
\langle\vec{J};\vec{J}\rangle_{\omega;c,t=1}\Big]+\left(dc+\frac{i\omega}{i(\omega+i\varepsilon)}\e^{-\varepsilon
c}\right) \e^{i\omega c}(\vec{J}(c);\vec{J})\,.
\end{align*}

If $\varepsilon\rightarrow 0, \langle \Gamma
\rangle\rightarrow\infty$, $d\rightarrow 0$, and
\[
M=\frac{\langle\dot{\vec{J}};\dot{\vec{J}}\rangle_{\omega;(c,\infty)}}
{\e^{i\omega
c}(\vec{J}(c),\vec{J})+\langle\dot{\vec{J}};\vec{J}\rangle_{\omega;(c,\infty)}}\,.
\]

This expression at small values of $c$ is close to
(\ref{m}). For the case \textit{3c)} with the distribution $p_{q}(u)$ of the  kind
(\ref{pqc}) the Laplace transform (\ref{la}) is substituted by
the opeation
\[
\langle A;B
\rangle_{\omega+i\varepsilon;(0,c)}+\e^{\textstyle c(\varepsilon_{2}-\varepsilon_{1})}\langle
A;B \rangle_{\omega+i\varepsilon_{2};(c,\infty)}\,,
\]
where $\langle
A;B\rangle_{\omega+i\varepsilon;(0,c)}={\textstyle\int\limits_{0}^{c}}\mathrm{d}t\,\e^{i(\omega+i\varepsilon)
t}(A(t),B)$,\\
$\langle A;B\rangle_{\omega+i\varepsilon_{2};(c,\infty)}$ is given
in (\ref{nla}) and the value $M$ from (\ref{m}) is replaced by
\[
M=\frac{\langle\dot{\vec{J}};\dot{\vec{J}}\rangle_{\omega+i\varepsilon_{1};(0,c)}+
{\textstyle\mathrm{e}}^{ (\varepsilon_{2}-\varepsilon_{1})}{\displaystyle\frac{i(\omega+i\varepsilon_{1})}{i(\omega+i\varepsilon_{2})}}
\langle\dot{\vec{J}};\dot{\vec{J}}\rangle_{\omega+i\varepsilon_{2};(c,\infty)}}{K_{2}}\,,
\]
\begin{align*}
K_{2}=(\vec{J}(0),\vec{J})-\e^{i(\omega+i\varepsilon_{1})c}(\vec{J}(c);\vec{J})+
\langle\dot{\vec{J}};\vec{J}\rangle_{\omega+i\varepsilon_{1};(0,c)}+ \\
\e^{c(\varepsilon_{2}-\varepsilon_{1})}\frac{i(\omega+i\varepsilon_{1})}{i(\omega+i\varepsilon_{2})}
\left[\e^{i(\omega+i\varepsilon_{2})c}(\vec{J}(c),\vec{J})+
\langle\dot{\vec{J}};\vec{J}\rangle_{\omega+i\varepsilon_{2};(c,\infty)}\right].
\end{align*}
At $\langle \Gamma \rangle\rightarrow\infty$ and
$\varepsilon_{2}\rightarrow0$ the value $M$ changes unessentially,
taking on the form
\[
M=\frac{\langle\dot{\vec{J}};\dot{\vec{J}}\rangle_{\omega+i\varepsilon_{1};(0,c)}+
{\displaystyle\e^{-c\varepsilon_{1}}\frac{i(\omega+i\varepsilon_{1})}{i\omega}}
\langle\dot{\vec{J}};\dot{\vec{J}}\rangle_{\omega;(c,\infty)}}{K_{3}}\,,
\]
\begin{align*}
K_{3}=(\vec{J}(0),\vec{J})-\e^{i(\omega+i\varepsilon_{1})c}(\vec{J}(c),\vec{J})+
\langle\dot{\vec{J}};\vec{J}\rangle_{\omega+i\varepsilon_{1};(0,c)}+ \\
\e^{-c\varepsilon_{1}}\frac{i(\omega+i\varepsilon_{1})}{i\omega}
\left[\e^{i\omega
c}(\vec{J}(c),\vec{J})+\langle\dot{\vec{J}};\vec{J}\rangle_{\omega;(c,\infty)}\right]\,.
\end{align*}
At small values of $\varepsilon_{1}$ this expression is
close to (\ref{m}). From (\ref{avc}) it is seen that
$\lim_{\varepsilon_{2}\rightarrow0}\varepsilon_{2}\langle \Gamma
\rangle=\e^{-\varepsilon_{1}c}$.

Let us summarise explicit results for Coulomb systems. Such
systems were investigated by the \emph{NSO} method in
\cite{zub97,rep88,Adams07,Rein,Adam}. We will follow \cite{zub97}.
We will derive the expressions for the conductivity of a
completely ionised Coulomb plasmas in a constant electric field.
For simplicity we limit ourselves to the case of the plasma
consisting of two subsystems, electrons and positive ions. An
isothermal limit is considered when the characteristic thermalization
time for the charge carriers is much less than the relaxation time
of their composite momentum. The formula for the isothermal
conductivity (at the frequency $\omega=0$) from \cite{zub97} is
written as follows:

\begin{equation}
\frac{1}{\sigma}=\frac{\beta}{3V}\left(\frac{m}{\e^{2}n}\right)^{2}\lim_{\varepsilon
\rightarrow
+0}\langle\dot{\vec{J}};\dot{\vec{J}}\rangle_{i\varepsilon},
\label{cond}
\end{equation}
where $V$ is system volume, $\beta=1/T$ is inverse temperature,
$m=m_{e}$ and $e$ are mass and charge of electron, $n=n_{e}$ is
average electron density. For the correlation function from
(\ref{cond}) in  an expression is obtained \cite{zub97}:

\begin{align*}
&\lim_{\varepsilon \rightarrow
+0}\langle\dot{\vec{J}};\dot{\vec{J}}\rangle_{i\varepsilon}= \\
&-\frac{1}{\beta}\left(\frac{e}{m}\right)^{2}\Sigma_{\vec{k}}\vec{k}^{2}v(\vec{k})S_{i}(\vec{k})\left[\lim_{\omega\rightarrow
0} \frac{1}{\omega}\mathrm{Im}\frac{1}{\epsilon_{e}(\vec{k},
\omega)}\right]\,,
\end{align*}
where $\vec{k}$ is wave vector, $v(\vec{k})=4\pi
\e^{2}/\vec{k}^{2}$, $S_{i}(\vec{k})$, the equilibrium
statistical structure factor of ions, $\epsilon_{e}(\vec{k},
\omega)$, equilibrium dielectric constant of an electronic
subsystem.

If we pass from the correlation function
$\langle\dot{\vec{J}};\dot{\vec{J}}\rangle_{i\varepsilon}$ of a
kind (\ref{la}) to the correlation function of a kind (\ref{la1})
in a case \textit{3a)} (\ref{pa}) at $\langle\Gamma\rangle
\rightarrow \infty, \varepsilon \rightarrow 0$ in (\ref{ava}) we
see that at $\omega=0$, ${\displaystyle
\frac{1}{\sigma}}={\displaystyle\frac{1}{\sigma}_{zub}\frac{a^{k}}{(a+c)^{k}}}$,
where ${\displaystyle \frac{1}{\sigma}_{zub}}$ is the expression
for the conductivity (\ref{cond}), derived in \cite{zub97}. For a
case \textit{3b)} with the distribution $p_{q}(u)$ (\ref{pqb}) at
$\langle\Gamma\rangle \rightarrow \infty, \varepsilon \rightarrow
0$, ${\displaystyle
\frac{1}{\sigma}}={\displaystyle\frac{1}{\sigma}_{zub}}$. For a
case \textit{3c)} with the distribution (\ref{pqc}) at
$\langle\Gamma\rangle \rightarrow \infty, \varepsilon_{2}
\rightarrow 0$ in (\ref{avc}), ${\displaystyle
\frac{1}{\sigma}}={\displaystyle\frac{1}{\sigma}_{zub}}
\e^{\textstyle -\varepsilon_{1}c}$.

Direct comparison of the obtained results to the experiment
presents certain difficulties, since the parameters
$c,a,k,\varepsilon_{1}$ are apriori unknown.

\section{Superstatistics from distribution containing lifetime}
 \label{sect:5}

In \cite{leont} it is pointed out that a nonequilibrium
distribution is characterized by an additional parameter related
to the deviation of a system from the equilibrium (caused by the field of
gravity, electric field for dielectrics etc). In the present work
we consider open nonequilibrium, stationary systems, certain point
of metastable states. Investigations on spin glasses and other
aging systems, where a "waiting time" plays an important role, allows to anticipate
the usability of this approach with respect to them as well. In the present paper we suggest a
new choice of an additional parameter in the form of the lifetime
of a physical system which is defined as a first-passage time till
the random process $y(t)$ describing the behaviour of the
macroscopic parameter of a system (energy, for example) reaches
its zero value. The lifetime $\Gamma_{x}$ (or $\Gamma$) is thus a
random process which is subordinate (in terms of the definitions
of the theory of random processes \cite{fe}) with respect to the
master process $y(t)$,
\[
\Gamma_{x}=\emph{inf}\{t: y(t)=0\}, \quad y(0)=x>0 \,.
\]
This definition of the lifetime is taken from the apparatus of the theory of random
processes where it is widely used in the theory of queues,
stochastic theory of storage \cite{prab}, Kramers problem of the
escape rate out of a potential well \cite{kr,me} and so on. These
questions are discussed in textbooks by van Kampen \cite{ka}, Gardiner
\cite{ga} and many other \cite{ta}. The lifetime plays part in
the theory of phase transitions, chemical reactions, in
the dynamics of complex biomolecules etc.

Using a maximum-entropy principle \cite{ja}, it is possible to
derive the form of the expression for microscopic (but coarse-grained)
probability density in the extended phase space
\cite{ry04,rys04,ry05,rys06}
\begin{equation}
\rho(z;E,\Gamma)=\exp\{-\beta
E-\gamma\Gamma\}/Z(\beta,\gamma)\,, \label{rho}
\end{equation}
where
\begin{align}
&Z(\beta,\gamma)= \int \exp\{-\beta
E-\gamma\Gamma\}\,\mathrm{d}z=
\label{z} \\
&\int\int
\mathrm{d}E\,\mathrm{d}\Gamma\omega(E,\Gamma)\exp\{-\beta
E-\gamma\Gamma\} \nonumber
\end{align}
is the partition function, $\beta$ and $\gamma$  are Lagrange
multipliers satisfying the equations for the averages
\begin{equation}
\langle E\rangle=- \frac{\partial \ln Z}{\partial\beta}
_{|\gamma} \, ; \quad  \langle\Gamma\rangle=-\frac{\partial
\ln Z}{\partial\gamma} _{|\beta}\,. \label{steq}
\end{equation}

The distribution (\ref{rho}) with the lifetime contains two
different time scales: the first relates to the energy $E$, and
the second - to the lifetime itself  $\Gamma$, this latter one
accounts for large-scale time correlations and large-time changes
in $E$ by means of a thermodynamic conjugate to the lifetime
value $\gamma$. The similar operation can be derived starting from
\emph{NSO}. The structure factor $\omega(E)$ is thus replaced by
$\omega(E,\Gamma)$ - the volume of the hyperspace containing given
values of $E$ and $\Gamma$. The number of phase points between
$\{ E,E+\mathrm{d}E$; $\Gamma,\Gamma+\mathrm{d}\Gamma\}$ equals
$\omega(E,\Gamma)\mathrm{d}E\,\mathrm{d}\Gamma$. The value thermodynamically
conjugated to the lifetime is related to the entropy fluxes and
entropy production which characterize the peculiarities of the
nonequilibrium processes in an open thermodynamic system. If
$\gamma=0$ and $\beta=\beta_{0}=1/(k_{B}T_{eq})$, where $k_{B}$ is
the Boltzmann constant, $T_{eq}$  is the equilibrium temperature,
then the expressions (\ref{rho})-(\ref{steq}) yield the
equilibrium Gibbs distribution. One can thus consider
(\ref{rho})-(\ref{steq}) as a generalization of the Gibbs
statistics towards the nonequilibrium situation. Such physical
phenomena as the metastability, phase transitions, stationary
nonequilibrium states are known to violate the equiprobability of
the phase space points. The value $\gamma$  can be regarded as a
measure of the deviation from the equiprobability hypothesis. In
general one might choose the value $\Gamma$ as a subprocess of
some other kind as chosen above. Mathematically the introduction of the
lifetime means acquiring additional information regarding an
underlying stochastic process; namely, exploring the (stationary) properties of
its slave process beyond merely knowledge of its
stationary distribution.

In the distribution (\ref{rho}) containing lifetime as a
thermodynamic parameter, the probability for $E$ and $\Gamma$ is
equal to
\begin{equation}
P(E,\Gamma)=\frac{\e^{-\beta
E-\gamma\Gamma}\omega(E,\Gamma)}{Z(\beta,\gamma)}. \label{P}
\end{equation}
Having integrated (\ref{P}) on $\Gamma$, we obtain the distribution
of a kind
\begin{equation}
P(E)=\int P(E,\Gamma)d\Gamma=\frac{\e^{-\beta
E}}{Z(\beta,\gamma)}\int_{o}^{\infty}\omega(E,\Gamma)\e^{-\gamma\Gamma}d\Gamma.
\label{pe}
\end{equation}

The structural factor $\omega(E,\Gamma)$  has a sense of the joint
probability for $E$ and  $\Gamma$, considered as a stationary
distribution of this process. We shall write down
\begin{equation}
\omega(E,\Gamma)=\omega(E)\omega_{1}(E,\Gamma)=\omega(E)\sum_{k=1}^{n}R_{k}f_{k}(\Gamma,E)\,.
\label{om}
\end{equation}
In the last equality (\ref{om}) it is supposed, that there exist $n$
classes of ergodic states in a system; $R_{k}$  is the probability
that the system is in the $k$-th class of ergodic states,
$f_{k}(\Gamma,E)$  is the density of lifetime distribution $\Gamma$
in this class of ergodic states (generally $f_{k}$ depends on
$E$). As a physical example for such situation (typical for
metals or glasses) one can mention the potential of many complex
systems. This case is considered in \cite{ol93}.

In \cite{ry05,rys06} various models of superstatistics are obtained
from (\ref{rho})-(\ref{om}). For example:
\[
P(E)\sim\frac{\e^{\textstyle -\beta
E}}{\left[1+(q-1)\left(1-\e^{\textstyle -\beta_{0}r_{0}E}\right)\right]^{\textstyle\frac{1}{q-1}}};
\]
\begin{align*}
& P(E)\sim \\
& \frac{\e^{-\beta
E}}{\left\{1+(q-1)(1-\left[1+(q_{1}-1)\left(1-\e^{\textstyle
-\beta_{0}r_{0}E}\right)\right]^{\textstyle\frac{-1}{q_{1}-1}})\right\}^{\textstyle\frac{1}{q-1}}},
\end{align*}
and so on.

We note that there is a similarity between the method of
superstatistics, where the averaging is performed over a parameter
$\beta$ (for example, the inverse temperature), as in (\ref{su}),
and the method of \emph{NSO}, where averaging is performed over
the extension of past time $u=t-t_{0}$, as in (\ref{NSO})
\cite{ryge}. Expressions (\ref{su}), (\ref{NSO}) and (\ref{16})
are described by the subordinated random processes \cite{fe}. The
Zubarev approach claims that the source term should be
infinitesimally small. The question is now whether a vanishing
source term would yield results different from the
superstatisrtics as well. In \cite{beck,beck1,beck01} a very
simple example  of the Brownian particle is considered. Its
velocity $v$ satisfies the linear Langevin equation
$\dot{v}=-\gamma v+\sigma L(t)$ where $L(t)$ is Gaussian white
noise, $\gamma > 0$ is friction constant, and the strength of the
noise is controlled in a usual fashion by the parameter $\sigma$.
The stationary probability density of $v$ is Gaussian with average
0 and variance $\beta^{-1}$, where the parameter
$\beta=\gamma/\sigma^{2}$ can be identified with the inverse
temperature of the statistical mechanics (we assume that the
Brownian particle has a unit mass). This simple situation
completely changes if the parameters $\gamma$ and $\sigma$ in the
stochastic differential equation are assumed to fluctuate as well.
To be specific, let us adopt that either $\gamma$ or $\sigma$ or
both fluctuate in such a way that $\beta=\gamma/\sigma^{2}$ is
$\chi^{2}$- distributed with degree $n$. This implies that the
probability density of $\beta$ is given by
\begin{align}
 & f(\beta)=\frac{1}{\Gamma(n/2)}\left(\frac{n}{2\beta_{0}}\right)^{\textstyle \frac{n}{2}}
\beta^{\textstyle \frac{n}{2}-1}\exp\left(-\frac{n\beta}{2\beta_{0}}\right) \,;  \nonumber \\
& \beta_{0}=\int_{0}^{\infty}\beta f(\beta) d \beta. \label{gdis}
\end{align}
In the Zubarev approach if the parameter $\varepsilon\rightarrow
0$ in the exponential distribution (\ref{expDi}), the source in
the Liouville equation vanishes. Relating to the distribution
(\ref{gdis}) it corresponds to $\beta_{0}\rightarrow\infty$ and
$(n/\beta_{0})\rightarrow 0$. As $u\leftrightarrow\beta$,
$p_{q}(u)\leftrightarrow f(\beta)$,
$\langle\Gamma\rangle\leftrightarrow\beta_{0}$ in (\ref{su}),
(\ref{NSO}), (\ref{gdis}), the case
$\langle\Gamma\rangle\rightarrow\infty$ corresponds to that
$\beta_{0}\rightarrow\infty$. Apparently, this is the limiting
case of $\sigma^{2}\rightarrow 0$, when no stochastic element is
present, but the system is subject to the dynamical force only, as
in the Liouville equation without random source.

\section{Piecewise
continuous distributions for functions $R, f$ from correlations
(\ref{pe})-(\ref{om}), (\ref{su})} \label{sect:6}

In this section the distributions of lifetime, having a different
shape on the different temporal intervals of evolution of the
system, are considered. Such behaviour is characteristic for
many physical systems. It is stressed in \cite{bog} that
non-equilibrium systems can have different stages of evolution. In
 \cite{hi} it is shown that the first passage time
probability density distribution changes depending on the value of
control parameter. Solutions of the Kramers equation, which are
related to the first passage time probability density distribution
also depend on the control parameter. Such transitions in real
systems are widely encountered, the aging of materials is just
one of known examples thereto.

Let us write the expressions (\ref{pe})-(\ref{om}) in the form
\begin{equation}
P(E)=\frac{\e^{-\beta
E}\omega(E)}{Z(\beta,\gamma)}\int\limits_{0}^{\infty}R(x)f_{1}(x,E)\mathrm{d}x\,;
\label{16}
\end{equation}
\[
f_{1}(x,E)=\int_{o}^{\infty}\e^{-\gamma\Gamma}f(x,E,\Gamma)d\Gamma\,.
\]
We note, that the correlation (\ref{16}) includes Laplace
transform to which  a probabilistic sense
can be ascribed according to \cite{kli}. For $f$ and $f_{1}$ from (\ref{16}) it is
possible to use the models \cite{ry04,rys04,ry05,rys06} which
leads to superstatistics of the kind $\exp\{-yE\}$. The similar
approach developed in \cite{str61} allows to obtain  the correlation value $\Gamma_{0}(y)\sim
a\e^{kyE}$ for the model
of phase synchronization, which under certain conditions reduces to the form
$\exp\{-yE\}$. The parameters $a$ and $k$ depend on the
problem.

Let us consider this problem for a simple case of the
function $R(y)$ from (\ref{16}), allowing to write an obvious
form of probability density where the function $R(y)$ represents a
combination of delta-function and homogeneous distribution:

\begin{align}
R(y)=
\left\{
\begin{array}{ll}
p\delta(y-a), \qquad y<c, \quad a<c, \quad p<1\,; &
\\
(1-p)m^{-1},  \quad 0<c<y<c+m\,; &
\\
0  \quad  y\geq m\,. &
\end{array}
\right. \nonumber
\end{align}

Then
\begin{align*}
& p(E)=\int\limits_{0}^{\infty}R(y)\frac{\,\mathrm{d}y}{Z_{1}(1+\gamma a\e^{kyE})}=
 \frac{1}{Z_{1}}\times \\
&\left[\frac{p}{(1+\gamma
a\e^{kyE})}+(1-p)\left(m+\frac{1}{kE}\ln\left|\frac{\displaystyle
\frac{1}{\gamma a}+\e^{kEc}}{\displaystyle \frac{1}{\gamma
a}+\e^{kE(c+m)}}\right|\right)\right]
\end{align*}

Let us choose now the expressions of a kind
\[
P(E)=\int\limits_{0}^{\infty}R(y)\frac{1}{Z}\e^{\textstyle-\kappa yE}\,\mathrm{d}y
\]
(coinciding with (\ref{su})) with piecewise continuous
distribution of the function $R(y)$. If we set the function $R(y)$
in the form of the gamma-distribution with different values of the
parameters $\alpha$ and $r$ in different areas,
\begin{equation*}
R(y)=\left\{
\begin{array}{l}
{\displaystyle \frac{\lambda^{\alpha}}{\Gamma(\alpha)}}y^{\alpha-1}\e^{-\lambda E}\,,
\quad y<c \,;
\\
b{\displaystyle\frac{\lambda^{r}}{\Gamma(r)}}y^{r-1}\e^{-\lambda E}, \quad y\geq
c,
\end{array}
\right.
\end{equation*}
then $b={\displaystyle\frac{\Gamma(r)\Gamma(\alpha,\lambda
c)}{\Gamma(\alpha)\Gamma(r,\lambda c)}}$, and
\begin{align}
p(E)=\frac{1}{Z\Gamma(\alpha)}\left[\gamma(\alpha,(\kappa
E+\lambda)c)\frac{1}{(1+\kappa E/\lambda)^{\alpha}} +  \right.  \label{19} \\
\left. \frac{\Gamma(\alpha,\lambda c)}{\Gamma(r,\lambda
c)}\Gamma(r,(\kappa E+\lambda)c)\frac{1}{(1+\kappa
E/\lambda)^{r}}\right]\,. \nonumber
\end{align}
This distribution is more complex, than Tsallis distribution
\cite{ts,abe}, obtained from gamma-distribution by means of a
method of superstatistics \cite{beck,beck1}. Multipliers in a form
$1/(1+\kappa E/\lambda)^{\alpha}$ correspond to the Tsallis
distribution, but in (\ref{19}) other factors depending on E are
present. The distribution (\ref{19}) passes in
$1/(1+\kappa E/\lambda)^{\alpha}$ at $r\rightarrow\alpha$.

If
\begin{align*}
R(y)=\left\{
\begin{array}{l}
{\displaystyle\frac{\lambda^{\alpha}}{\Gamma(\alpha)}}y^{\alpha-1}\e^{-\lambda E},
\quad y<c ;
\\
b{\displaystyle\frac{ka^{k}}{(y+a)^{k+1}}}, \quad y\geq c\,,
\end{array}
\right.
\end{align*}
then  $b={\displaystyle \frac{k(a+c)^{k}\Gamma(\alpha,\lambda
c)}{\Gamma(\alpha)}}$, and
\begin{align*}
p(E)=\frac{1}{Z\Gamma(\alpha)}\left[\gamma(\alpha,(\kappa
E+\lambda)c)\frac{1}{(1+\kappa E/\lambda)^{\alpha}} + \right. \\
 k(a+c)^{k}\Gamma(\alpha,\lambda c)\e^{a\kappa E}(\kappa
E)^{k}\Gamma(-k,\kappa E(c+a))\Big]\,.
\end{align*}
Other similar examples can be
considered as well. One could set three and more areas of
piecewise change of variables (for example, two areas
corresponding to different phases and a transitive layer between
them). A continuous change of parameters can be treated on equal
footing, for example, considering the parameter of the
gamma-distribution continuously changing with some distribution
function or to be set by functions of a kind $R(g(\cdot))$. A
combination of discrete $R_{k}$ and continuous $R(y)$
distributions in different areas is also possible.

Let us consider two more simple examples of the combination of
gamma-distribution with delta-distribution and with homogeneous
distribution:
\begin{equation*}
R(y)=\left\{
\begin{array}{l}
{\displaystyle\frac{\lambda^{\alpha}}{\Gamma(\alpha)}}y^{\alpha-1}\e^{-\lambda E}\,,
\quad y<c \, ;
\\
b\,\delta(y-d)\,, \quad y\geq c \,, d>c\,.
\end{array}
\right.
\end{equation*}
Then $b={\displaystyle \frac{\Gamma(\alpha,\lambda c)}{\Gamma(\alpha)}}$,
\begin{align*}
p(E)=\frac{1}{Z\Gamma(\alpha)}\Big[\gamma(\alpha,(\kappa
E+\lambda)c)\frac{1}{(1+\kappa
E/\lambda)^{\alpha}}+\\
\Gamma(\alpha,\lambda c)\e^{-d\kappa E}\Big]\,.
\end{align*}

By

\begin{align}
R(y)= \left\{
\begin{array}{ll}
{\displaystyle\frac{\lambda^{\alpha}}{\Gamma(\alpha)}}y^{\alpha-1}\e^{-\lambda E}\,,
\quad y<c\,; &
\\
bm^{-1}, \quad 0<c<y<c+m\,; &
\\
0,  \quad  y\geq m\,, &
\end{array}
\right. \nonumber
\end{align}

\[b={\displaystyle\frac{\Gamma(\alpha,\lambda c)}{\Gamma(\alpha )}}\,,\]

\begin{align*}
p(E)=\frac{1}{Z\Gamma(\alpha)}\Big[\gamma(\alpha,(\kappa
E+\lambda)c)\frac{1}{(1+\kappa
E/\lambda)^{\alpha}}+ \\
\Gamma(\alpha,\lambda c)\e^{-c\kappa E}(\kappa
E)^{-1}(1-\e^{-\kappa Em})\Big]\,.
\end{align*}

Various combinations of functions of distribution from
\cite{beck,beck1} can be brought into consideration: for example, lognormal
superstatistics, gamma superstatistics and inverse gamma
superstatistics.

\section{Applications} \label{sect:7}

Passing to superstatistics we get different distributions $R(y)$
from (\ref{16}) at $y<c$ and $y>c$ or $f(\beta)$ from (\ref{su})
at $\beta<c$ and $\beta>c$. For example, exponential
or gamma-distribution at $\beta<c$ and Pareto distribution
corresponding to Tsallis distribution at $\beta>c$.  A multitude of
various combinations of different distributions can be encountered, including,
for example, gamma-distribution with various parameters at different
temperatures, the subordinated distributions and so on. The parameters of
gamma-distribution can change discretely, but can be continuous as
well.

In \cite{abu} it is assumed that Beck and Cohen's superstatistics
provides a suitable description for systems with mixed
regular-chaotic dynamics. Such systems can be described by means of
the suggested approach. Examples of different behaviour of systems
at different temperatures are obvious including superconductivity,
and superfluidity, and other phase transitions.

The approach which is more general, than the superstatistics
theory, consists in setting  piecewise continuous distributions
for $R(y)$ from expression (\ref{16}), that is the probabilities
for a system to be in in the $k$-th state.

Such distributions can describe laminar and
turbulent modes in the stream, for example, the diffusion of the
tobacco smoke flow in the atmosphere. For this case the index $k$
(the parameter $y$ - in a continuous case) corresponds to the
spatial coordinate of the flow. In any point $c$ there is a
transition from a laminar mode to turbulent. The distribution
$R(y)$ or $f(\beta)$ can be described by the correlations obtained
in \cite{be7} (lognormal distribution). The situation described
here is more general, than the only one transition at a certain
temperature, as in the case of superstatistics.

In \cite{has} the entropic index can take various values, as in
the present work. In \cite{flei} the entropic parameter $q$ of the
Tsallis distribution depend on the parameter $m$, and the
parameter $m$ is used to give account of an energy loss rate or
energy dissipation rate (or perhaps, the energy absorption rate).
The Tsallis distribution appears to depend on the parameter $m$.
Such situation corresponds to the approach of the present work.

The distribution suggested in \cite{tss} represents a special case
of the piecewise continuous distribution used in the present work.
In \cite{tset} the problems close to the problems of the present
work are considered: the occupation of the accessible phase space
(or of a symmetry-determined nonzero-measure part of it), which in
turn appears to determine the entropic form to be used.

The results similar to results of Section 5 are obtained in
\cite{vak}. For example, the distribution $f(\rho_1 \dots \rho_N)$ of\\
metastable states with local fluid densities $\rho_{i}$ in
different spatial domains $i = 1\dots N$ depending from the
exponent $\lambda$ is again related to the distance from the
conventional equilibrium, as does the value $\gamma$  from Section 5.

Besides the spatial heterogeneity the piecewise continuous
distribution can describe the time changes. The suggested approach
allows to use the methods of the theory of random processes for
treating specific problems; for example, to refer to the
stochastic theory of storage \cite{prab,sp}, setting rates of an
input in a potential well and an output from it, and to the
methods of the Kramers escape problem \cite{kr,me}.

\section{Conclusion}
 \label{sect:8}

As it is stated in \cite{rau}, the existence of different time
scales and the flow of the information from slow to fast degrees
of freedom create the irreversibility of the macroscopical
description. The information thus is not lost, but passes in the
form inaccessible at the Markovian level of description. For
example, for the rarefied gas the information is transferred from
one-particle observable to multipartial correlations. In
\cite{ry01} the values  $\varepsilon=1/\langle\Gamma\rangle$ and
$p_{q}(u)=\varepsilon \exp\{-\varepsilon u\}$ are expressed
through the operator of entropy production and, according to the
results of \cite{rau}, in terms of the flow of the information
from relevant to irrelevant degrees of freedom. The introduction
of the function $p_{q}(u)$ in \emph{NSO} corresponds to the
specification of the description by means of the effective account
of communication with irrelevant degrees of freedom. In the
present work it is shown, how it is possible to expand the
specification of the description of memory effects within the
limits of \emph{NSO} method. A more detailed description of the
influence of fast varying variables on the evolution of system is
suggested based on specifying the density of the life span
distribution of a system.

In many physical problems the finiteness of the lifetime can be
neglected. Then
$\varepsilon\sim 1/\langle\Gamma\rangle\rightarrow 0$. For example,
for a case of the evaporation of liquid drops it is possible to
show \cite{ryDr}, that non-equilibrium characteristics depend on
$\exp\{y^{2}\}$; $y=\varepsilon/(2\lambda_{2})^{1/2}$, $\lambda_{2}$
is the second moment of the correlation function of fluxes
averaged over quasi-equilibrium distribution. Estimations show,
that even at the minimal values of lifetime of drops (generally of
finite size) the maximum sizes are
$y=\varepsilon/(2\lambda_{2})^{1/2}\leq 10^{-5}$. Therefore
finiteness of values of $\langle\Gamma\rangle$ and $\varepsilon$ does
not influence the behaviour of system and it is possible to
consider $\varepsilon=0$. However in some situations it is
necessary to take into account the finiteness of lifetime
$\langle\Gamma\rangle$ and values $\varepsilon> 0$. For example,
this is the case of the nano-drops.

Changes of the form of the source in the Liouville equation, as
well as the expressions for the kinetic coefficients, average
fluxes, and kinetic equations can be obtained with the use of the
\emph{NSO}. It is possible to choose a class of lifetime
distributions for which after thermodynamic limiting transition
and tending the average lifetime of system to infinity the results
are reduced to those obtained under exponential distribution for
lifetime, used by Zubarev. However there is also another extensive
class of realistic distributions of lifetime of system for which
even if the average lifetime of system tends to infinity the
non-equilibrium properties essentially change. It is a consequence
of the interaction of the system with its environment.

For the distributions of the kind (\ref{pa}), (\ref{pqb}),
(\ref{pqc}), having different form for different argument spans,
non-vanishing corrections to Zubarev \emph{NSO} persist even for
infinitely large systems with infinitely large lifetimes. In the
present work it is shown, that this situation is possible, for
example, for the distributions of lifetime of a system, having
different form at different stages of the evolution of a system.
Such behaviour corresponds to the realization of the evolution of
a system in a number of subsequent stages \cite{bog}.

It is interesting to investigate possibilities of such a choice of the
function $p_{q}(u)$ which would in most full fashion correspond to physical
conditions in which a system is placed. For an optimal choice of a
form of function $p_{q}(u)$ it is possible to use the method of
maximum entropy principle.

The use of the theory of the superstatistics in various
applications is related to the piecewise continuous functions for
the density of distribution of lifetime of the system. The task of
these functions of distribution allows to receive new expressions
for the superstatistics, corresponding to various physical
situations. Similar results seem to be useful, for example, in the
investigation of small systems. A number of results following
from the interpretation of \emph{NSO} and $p_{q}(u)$ as a density of
lifetime distribution of system \cite{ry01}, can be obtained from
the stochastic theory of storage \cite{prab} and theories of
queues. For example, in \cite{prab} the general result that the
random variable of the period of employment (lifetime) $g(u,x)$ has
absolutely continuous distribution $p_{q}(u)\equiv g(u,x)=xk(u-x,u)$,
$u>x>0$, and $g(u,x)=0$ otherwise, is stated. There $k(x,t)$ is an
absolutely continuous distribution for the value $X(t)$ of the input
into a system.

Pursuing the analogy between the methods of \emph{NSO} and the
superstatistics, it is necessary to consider in more detail a case
of dynamical representation of the superstatistics when
$\beta_{0}\rightarrow\infty$, and the distribution of an intensive
parameter has a piecewise continuous character.

The form of distribution chosen by Zubarev for the life span
represents a certain limiting case. The choice of the lifetime
distribution in \emph{NSO} is related to the account of the past
of a system, its physical features, on the present moment; for
example, with the account of the age of a system only, as in
Zubarev form of \emph{NSO} \cite{zub71,zub80,zub96,ry01,ra95,ra99}
at $\varepsilon>0$, or with more detailed characteristic of the
past evolution of a system. The obtained results are essential in
cases when it is impossible to neglect the memory effects since
the memory correlation time there is not vanishing. The analysis
of the corresponding time scales is necessary as it is noted in
\cite{rau}.

The main objective of the present article is to show, how the systems
with infinitely large average lifetime can induce nonvanishing sources
in the Liouville equation, and in what consequences for the  method of
the \emph{NSO} it results. Superstatistics with piecewise
continuous distributions of intensive parameter are considered
as well.

\end{document}